\documentclass[a4paper,twocolumn,11pt,accepted=2021-08-01]{quantumarticle}
\pdfoutput=1
\usepackage[utf8]{inputenc}
\usepackage[english]{babel}
\usepackage[T1]{fontenc}
\usepackage{amsfonts,amssymb,amsmath}
\usepackage{hyperref}

\usepackage{tikz}
\usepackage{lipsum}

\begin{document}

\title{Borromean states in discrete-time quantum walks}

\author{Marcin Markiewicz}   
\affiliation{International Centre for Theory of Quantum Technologies (ICTQT), University of Gdansk, 80-308 Gdansk, Poland}

\author{Marcin Karczewski}   
\affiliation{International Centre for Theory of Quantum Technologies (ICTQT), University of Gdansk, 80-308 Gdansk, Poland}

\author{Pawe\l{} Kurzy\'nski}   
\affiliation{Institute of Spintronics and Quantum Information, Faculty of Physics, Adam Mickiewicz University in Pozna{\'n}, Uniwersytetu Pozna{\'n}skiego 2, 61-614 Pozna\'n, Poland}



\maketitle 

\begin{abstract}

In the right conditions, removing one particle from a multipartite bound state can make it fall apart. This feature, known as the "Borromean property", has been recently demonstrated experimentally in Efimov states. One could expect that such peculiar behavior should be linked with the presence of strong inter-particle correlations. However, any exploration of this connection is hindered by the complexity of the physical systems exhibiting the Borromean property. To overcome this problem, we introduce a simple dynamical toy model based on a discrete-time quantum walk of many interacting particles. We show that the particles described by it need to exhibit the Greenberger-Horne-Zeillinger (GHZ) entanglement to form Borromean bound states. As this type of entanglement is very prone to particle losses, our work demonstrates an intuitive link between correlations and Borromean properties of the system. Moreover, we discuss our findings in the context of the formation of composite particles.

\end{abstract} 


\section{Introduction}

\begin{figure}[t]
\centering
\includegraphics[scale=1]{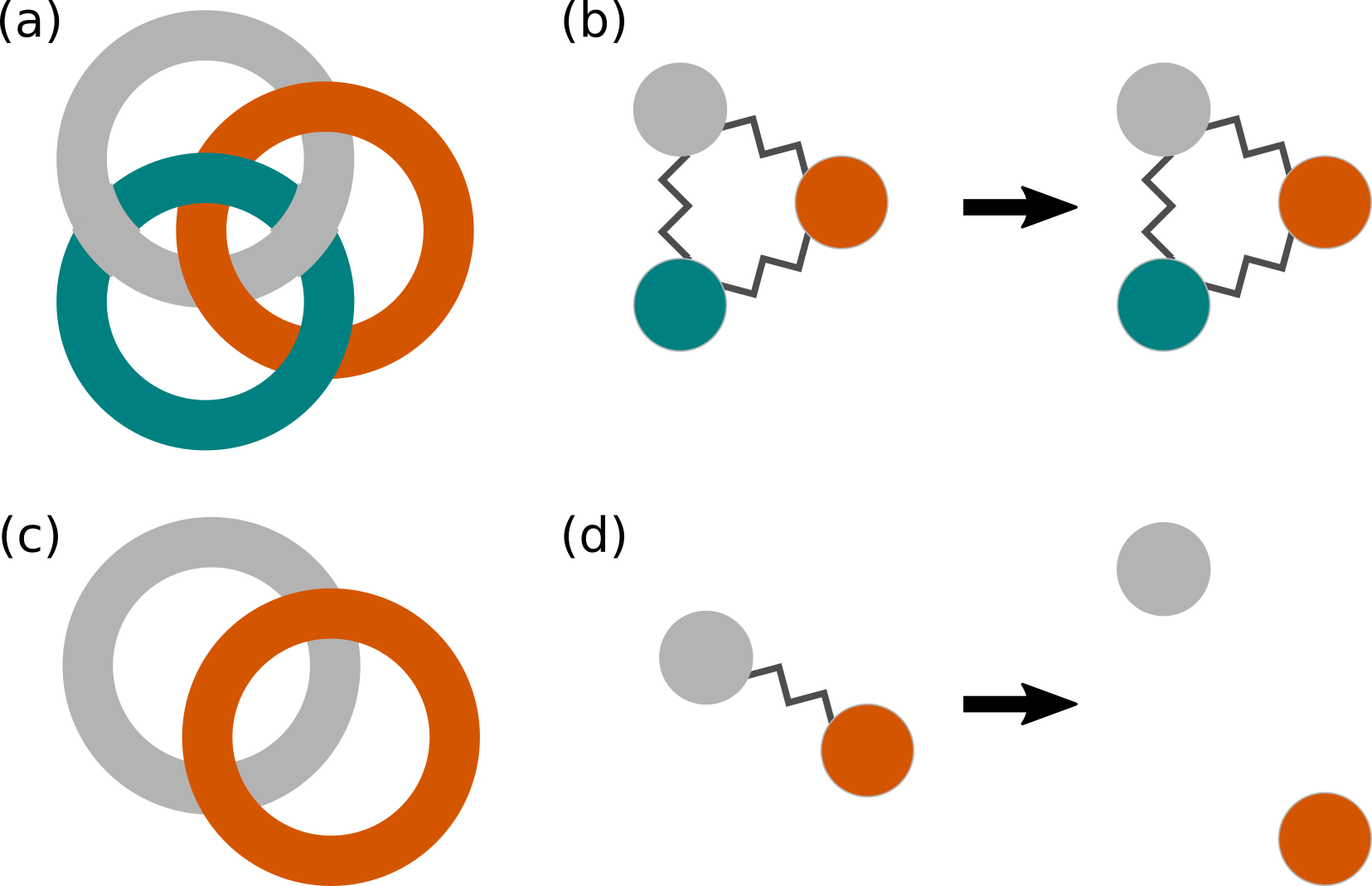}
\caption{To disconnect Borromean rings (a), one of them needs to be cut apart (c). Borromean compound (b) breaks down when a particle is removed (d).}
\label{fig1}
\vspace{-1mm}
\end{figure}

Borromean state is a bound state of three quantum particles that falls apart if one particle is removed. Its name originates from Borromean rings, a peculiar construction that holds together due to a genuine tripartite arrangement, see Fig. \ref{fig1}. No two elements are directly connected -- it requires a triple to make a stable structure. A generalization of this concept to $n$ elements is known as an $n$-component Brunnian link \cite{brunnian,brunnian2}.

The physical origin of Borromean states is quite counterintuitive. One would expect that if pairwise interactions enable a multipartite compound to arise, they should also keep together what is left after a particle is removed from it.
However, in some situations interactions may not allow for the formation of bipartite bound states (dimers), but at the same time lead to the emergence of tripartite bound states (trimers). The most known example is the Efimov state \cite{Efimov}, a bound state of three identical bosons that, apart from its Borromean binding, exhibits scale invariance in a sense that there exists an infinite sequence of tripartite bound states whose scattering lengths and energies follow a geometric progression.

Even though Borromean states are believed to occur in many areas of quantum physics, it is not easy to observe them in natural conditions. It took 35 years to experimentally verify the existence of Efimov states in an ultracold gas of caesium atoms \cite{EfimovExperiment} (for a review see e.g. \cite{EfimovReview}). It is therefore important to  better understand the mechanism behind the formation of Borromean binding. To identify its essential properties, the emergence of  Borromean states should be studied  in simple dynamical systems, the simpler the better. 

In this work we investigate Borromean states in discrete-time quantum walks (DTQWs) \cite{Aharonov,Meyer}, a basic model of quantum particle-dynamics that is known to simulate a broad range of physical phenomena and was already implemented on many experimental platforms (for a review see \cite{Review1,Review2,Review3,Review4}). To this end, we allow the particles in one-dimensional DTQW to interact and form bound states. In general, interaction makes quantum walks hard to treat analytically even in the case of only two particles \cite{IQW1,IQW2,IQW3}. However, our model features Borromean states that can be determined exactly. 

Moreover, it allows us to investigate their properties from the quantum information perspective. Interestingly, it turns out that the Borromean states emerging in our model need to exhibit Greenberger-Horne-Zeilinger (GHZ) \cite{GHZ} type of entanglement between the internal degrees of freedom of the particles. Such entanglement is genuinely multipartite and is very sensitive to particle losses, which can be considered a Borromean property \cite{GHZBorr, GHZBorr2}. This showcases an intuitive link between the properties of particles' correlations and their dynamics. We further explore this connection from the point of view of formation of multipartite composite bosons \cite{Law,Us}.

\section{Multipartite interacting DTQW.} 

We consider a quantum walk describing the movement of $N$ distinguishable particles on a one-dimensional lattice (see Appendix \ref{app:dtqw}). Their positions are denoted as $\pmb{\mathrm{x}}=(x_1,\ldots,x_N)$, where $x_i\in \mathbb Z$. The particles move left or right according to their internal degree of freedom $c_i\in\{{\leftarrow},{\rightarrow}\}$, called coin. The total state of such a system is then:
\begin{eqnarray}
|\psi\rangle &=&\sum_{x_i,c_i}\alpha_{x_1,\ldots,x_N,c_1,\ldots,c_N}|x_1,c_1\rangle\otimes\ldots\otimes|x_N,c_N\rangle\nonumber\\
&\equiv& \sum_{\pmb{\mathrm{x,c}}}\alpha_{\pmb{\mathrm{x,c}}}|\pmb{\mathrm{x,c}}\rangle.
\end{eqnarray}
Its unitary evolution proceeds in discrete steps 
\begin{equation}\label{evo}|\psi_{t+1}\rangle=U_N|\psi_t\rangle=S^{\otimes N} C(\pmb{\mathrm{x}})|\psi_t\rangle,
\end{equation}
where $S$ is a conditional translation that shifts the position of each particle according to its coin 
\begin{equation}
S|x_i,{\rightarrow}\rangle = |x_i+1,{\rightarrow}\rangle,~~S|x_i,{\leftarrow}\rangle = |x_i-1,{\leftarrow}\rangle
\end{equation}
and $C$ encodes the bipartite interaction present in our model. We choose an interaction model in which the only moment $C$ is different from identity is when two, or more, particles meet. Intuitively this means that particles travel in one direction without dispersion until they meet. Only then the interaction occurs and the direction of their  movement changes. Physically this may correspond to a dense quantum gas in which the average time between particle collisions is much shorter than the time it takes a single-particle wave packet to disperse. Moreover, many classical cellular automata that simulate multipartite scatterings are based on the same assumption \cite{CA}. 

For any pair of particles the action of $C$ is given by 
\begin{equation}\label{C}
C(x_i,x_j)= \left\{\begin{matrix}
\openone_c \otimes \openone_c ~~\text{if}~~ x_i\neq x_j, \\
G_{ij}(\varphi)~~~\text{else}~~~~~~~~ 
\end{matrix}\right.,
\end{equation}
where $\openone_c$ is the identity on the coin space and
\begin{equation}\label{grov}
G_{ij}(\varphi) = \openone_c \otimes\openone_c  +(e^{i\varphi}-1)|+\rangle\langle +|\otimes |+\rangle\langle +|.
\end{equation}
In the above $|+\rangle = (|{\rightarrow}\rangle + |{\leftarrow}\rangle)/\sqrt{2}$ and the subspaces in tensor products correspond to the $i^{\mathrm{th}}$ and $j^{\mathrm{th}}$ particle. Notice that the operator $G_{ij}(\varphi)$ is a generalized Grover operator \cite{Grover}, whose standard form corresponds to $\varphi = \pi$. From now on we assume that $\varphi\neq0$, as in that case $G_{ij}$ is trivial and cannot lead to any Borromean states.  Finally, we would like to notice that the operator (\ref{C}) can be modified to include more complex dynamics when the particles are apart. For example, for $x_i\neq x_j$ one could choose $C(x_i,x_j)=H\otimes H$, where $H$ is the Hadamard operator. We will briefly discuss such a possibility at the end of this work.

Applying the definition (\ref{C}) to a multipartite case we see that if $N$ particles share the same position, the operator $C(\pmb{\mathrm{x}})$   implements a sequence of $\frac{N(N-1)}{2}$ generalized Grover operations. As all these operations commute, their order does not matter. For instance, if three particles indexed by $i,\,j$ and $k$ are at the same position, the total interaction is a product of three bipartite terms $G_{ij}(\varphi)G_{ik}(\varphi)G_{jk}(\varphi)$. 



\section{Borromean states.} 


We are going to look for bound states of N particles. Such states are often associated with  low energies. However, we cannot rely on this intuition, as the evolution of our system is discrete and the energies are defined only up to the multiples of some constant. Instead, we will use the following dynamical criterion: the particles forming a bound state remain \emph{close} to each other. Note that in the case of the interaction (\ref{C}) this means that such particles should  move collectively in one direction -- otherwise they would simply drift apart.  Moreover, they should also share the same position as we want the molecular binding to originate from the interaction, which in our model occurs only when the particles meet. Such a state is of the form:
\begin{equation}\label{general}
\sum_{x=0}^{d-1} |x,\ldots,x\rangle \otimes (\beta_x|{\rightarrow},\ldots,{\rightarrow}\rangle + \gamma_x|{\leftarrow},\ldots,{\leftarrow}\rangle),
\end{equation}
in which the coin subspace is in an N-partite GHZ state. In the above we assumed $x=0,1,\ldots,d-1$ and periodic boundary conditions. It should be stressed that the relatively simple form of the Borromean states (\ref{general}) is a feature of our model. In general, particles in such states do not need to share the same position or move in the same direction.

Since the evolution operator $U_N$ commutes with translation, we can assume that the state (\ref{general})  is its eigenvector. This means that  $\beta_x = \frac{e^{i\frac{2\pi}{d}kx}}{\sqrt{d}}\beta$ and $\gamma_x = \frac{e^{i\frac{2\pi}{d}kx}}{\sqrt{d}} \gamma$, where $k=0,1,\ldots,d-1$. Therefore, the sought bound eigenstates, if present, must be of the form
\begin{eqnarray}\label{eigenstate}
|\chi_k\rangle = \frac{1}{\sqrt{d}}\left(\sum_{x=0}^{d-1} e^{i\frac{2\pi}{d}kx}|x,\ldots,x\rangle \right) \otimes  |GHZ_N\rangle,
\end{eqnarray} 
where $|GHZ_N\rangle = \beta|{\rightarrow},\ldots,{\rightarrow}\rangle + \gamma|{\leftarrow},\ldots,{\leftarrow}\rangle$. Note that (\ref{eigenstate}) should satisfy
\begin{equation}
|\langle \chi_k|U_N|\chi_k\rangle | =1. 
\end{equation}
Using the definitions (2)-(5) and eq. (7) the above becomes
\begin{equation}\label{condition}
\left|\langle GHZ_N|P_k\left(\prod_{j<l}G_{jl}(\varphi)\right)|GHZ_N\rangle\right| = 1,
\end{equation}
where $\prod_{j<l}G_{jl}(\varphi)$ is a sequence of generalized Grover operators for all $\frac{N(N-1)}{2}$ pairs of particles and
\begin{eqnarray}
P_k &=& e^{-i\frac{2\pi k}{d}}|{\rightarrow},\ldots,{\rightarrow}\rangle\langle {\rightarrow},\ldots,{\rightarrow}| \nonumber \\
&+& e^{i\frac{2\pi k}{d}}|{\leftarrow},\ldots,{\leftarrow}\rangle\langle {\leftarrow},\ldots,{\leftarrow} |.
\end{eqnarray}
A solution to (\ref{condition}) exists only for $N=2,3,4$ and $k=0,d/2$ (for the details see Appendix \ref{app:borro}). More precisely, for $N=2$ the solution exists for an arbitrary $\varphi$ and $\beta=-\gamma$. For $N=3$ the solution exists for $\varphi=\frac{2\pi}{3}$ and $\beta=\gamma$. Finally, for $N=4$ the solution exists for $\varphi=\frac{2\pi}{3}$ and $\beta=-\gamma$. 

We fix $\varphi=\frac{2\pi}{3}$ and define
\begin{eqnarray}
|dim_r\rangle &=& \frac{1}{\sqrt{d}} \left( \sum_{x=0}^{d-1} (-1)^{rx}|x,x\rangle \right) \otimes |GHZ_2^-\rangle, \label{dim}\\
|tri_r\rangle &=& \frac{1}{\sqrt{d}} \left( \sum_{x=0}^{d-1} (-1)^{rx}|x,x,x\rangle \right) \otimes |GHZ_3^+\rangle, \label{tri} \\
|qua_r\rangle &=& \frac{1}{\sqrt{d}} \left( \sum_{x=0}^{d-1} (-1)^{rx}|x,x,x,x\rangle \right) \otimes |GHZ_4^-\rangle, \label{qua}\nonumber\\
\end{eqnarray}
where $|GHZ_N^{\pm}\rangle = \frac{1}{\sqrt{2}}(|{\rightarrow},\ldots,{\rightarrow}\rangle \pm |{\leftarrow},\ldots,{\leftarrow}\rangle)$ and $r=0,1$. Each of these six symmetric states is a product of two GHZ states: one of qudits from the position space and one of qubits from the coin space. They could describe the states of two, three, and four bosons. Finally, they are bound states, since all particles stay together, though their center of mass is completely delocalized. 

To prove the Borromean properties of $|tri_r\rangle$ and the Brunnian ones of $|qua_r\rangle$, we still need to show that they are no longer bound after a particle is removed from them. The resulting mixtures
\begin{equation}
\rho^{tri}_2 = \frac{1}{2d}\sum_{x=0}^{d-1}\sum_{c={\leftarrow},{\rightarrow}} |x,x,c,c\rangle\langle x,x,c,c|
\end{equation}
and
\begin{equation}
\rho^{qua}_3 = \frac{1}{2d}\sum_{x=0}^{d-1}\sum_{c={\leftarrow},{\rightarrow}} |x,x,x,c,c,c\rangle\langle x,x,x,c,c,c|,\end{equation}
 are clearly not the eigenstates of  $U_2$ and $U_3$. Moreover, we will show below that they are not bound, as the particles split during the evolution. 

A crucial observation is that once the coin operator produces a state in which one of two (or one of three) particles goes in a different direction, say $|{\rightarrow},{\leftarrow}\rangle$   (or $|{\rightarrow},{\rightarrow},{\leftarrow}\rangle$), the group splits during the conditional translation and will never reunite (in the $d\rightarrow\infty$ limit). In other words, once the particles leave the subspace given by the projector $\Pi_2 = 2d \,\rho_2^{tri}$  (or $\Pi_3 = 2d\, \rho_3^{qua}$), they cannot return. As a result, the relevant evolution operators are $V_2 = \Pi_2 U_2$ and $V_3 = \Pi_3 U_3$. These operators are contracting, i.e., the moduli of their eigenvalues $\lambda$ are less or equal one (for details, see Appendix \ref{app:two}). To check whether the particles stay together, we need to find the overlap between  $\rho_2^{tri}$ (or $\rho_3^{qua}$) and the eigenvectors for which $|\lambda |=1$. This is straightforward since we already proved that the only eigenvectors that satisfy the above condition are (\ref{dim}) and (\ref{tri}). As $\langle dim_r|\rho_2^{tri}|dim_r\rangle  = \frac{1}{2d}=\langle tri_r|\rho_3^{qua}|tri_r\rangle$, the particles in $\rho_2^{tri}$ and  $\rho_3^{qua}$ must split during the evolution.


\section{Deviation from ideal setting.} 

\begin{figure}[t]
\centering
\includegraphics[scale=0.65]{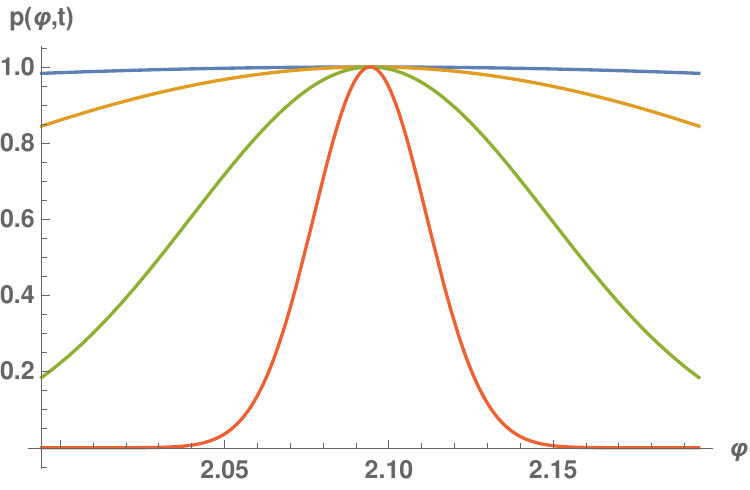}
\caption{Plot of $p(\varphi,t)$ for $t=1$ (blue), $t=10$ (orange), $t=100$ (green) and $t=1000$ (red).}
\label{fig2}
\end{figure}

The above Borromean and Brunnian states exist only for $\varphi =\frac{2\pi}{3}$. Let us investigate what happens if we deviate from this value. In particular, we consider the probability that the initial state $ |tri_0\rangle$ remains unchanged after $t$ steps $p(\varphi,t)=|\langle tri_0|V_3^t|tri_0\rangle|^2$. Since the position degrees of freedom can be factored out, the problem reduces to the evaluation of a $2\times 2$ expression 
\begin{eqnarray}
p(\varphi,t) &=& \left|\begin{pmatrix}1/\sqrt{2} & 1/\sqrt{2} \end{pmatrix} \begin{pmatrix} h & g \\ g & h \end{pmatrix}^t \begin{pmatrix} 1/\sqrt{2} \\ 1/\sqrt{2} \end{pmatrix}\right|^2\nonumber\\
&=&\left|\frac{e^{i3\varphi}+3}{4}\right|^{2t}, \label{fidelity}
\end{eqnarray}
where $g=\frac{1}{8}\left(-1+e^{i\varphi}\right)^2\left(2+e^{i\varphi}\right)$ and $h=\frac{1}{8}\left(4+3e^{i\varphi}+e^{i3\varphi}\right)$. The corresponding two-dimensional space is spanned by the coin states $|{\rightarrow},{\rightarrow},{\rightarrow}\rangle$ and $|{\leftarrow},{\leftarrow},{\leftarrow}\rangle$. The above implies that for small perturbation the state $|tri_0\rangle$ can still persist for many steps of the quantum walk, see Fig. \ref{fig2}.

\section{Borromean composite bosons.} 

As already mentioned, symmetric states (\ref{tri}) and (\ref{qua}) may describe a bound state of three and four identical bosons, respectively. Notice that as the evolution operator $U_N$ is also symmetric, our model can be applied to bosonic systems. Below we show that because of their GHZ entanglement, these states fulfill a necessary condition for describing a single composite boson  introduced in \cite{Law}. This complements the results obtained in \cite{Us}, where it was proven that $2N$ fermions must be genuinely multipartite entangled to behave as a single composite boson. Moreover, we highlight another Borromean property -- if a single elementary boson is removed from a composite particle, the compound loses its bosonic character. We focus on the state $|tri_0\rangle$, but the results for the other states are analogous.

Let $a^{\dagger}_{x,c}$ create a boson at position $x$ with a coin state $c$. The state $|tri_0\rangle$ can be rewritten as 
\begin{eqnarray}
|tri_0 \rangle = \frac{1}{\sqrt{3!}} \frac{1}{\sqrt{2d}} \sum_{x=1}^d \sum_{c=\leftarrow,\rightarrow}a^{\dagger 3}_{x,c} |\mathbf{0}\rangle, 
\end{eqnarray}
where $|\mathbf{0}\rangle$ is the vacuum state. Next, define an operator
\begin{eqnarray}
b^{\dagger} = \frac{1}{\sqrt{3!}} \frac{1}{\sqrt{2d}} \sum_{x=1}^d \sum_{c=\leftarrow,\rightarrow} a^{\dagger 3}_{x,c},
\end{eqnarray}
which creates a single composite particle with the internal structure formed by the positions and coin states of individual bosons. In other words, $b^{\dagger}|\mathbf{0}\rangle = |\mathbf{1}\rangle = |tri_0\rangle$.

Let us also introduce a state of $N$ composite particles
\begin{equation}
|\mathbf{N}\rangle = \frac{b^{\dagger N}}{\sqrt{B_N N!}}|\mathbf{0}\rangle,
\end{equation}
where $B_N$ is a normalization constant that reflects a deviation from a perfect bosonic behavior. It was proven in \cite{Law} that if $\frac{B_N}{B_{N-1}}\rightarrow 1$, creation and annihilation operators of a composite particle reproduce bosonic ladder structure, i.e., $b^{\dagger}|\mathbf{N}\rangle = \sqrt{N+1}|\mathbf{N+1}\rangle$ and $b|\mathbf{N}\rangle = \sqrt{N}|\mathbf{N-1}\rangle$. In particular, if $b^{\dagger}$ creates a perfect composite boson,  we need to have $\frac{B_2}{B_1} = B_2 = 1$ (note that by definition $B_1=1$). Therefore, our goal is to find 
\begin{equation}
B_2 = \frac{1}{2}\langle \mathbf{0} |b^2 b^{\dagger 2}|\mathbf{0}\rangle. 
\end{equation}
By evaluating the norm of $b^{\dagger 2}|\mathbf{0}\rangle$ (see Appendix \ref{app:comp}) we obtain 
\begin{equation}
B_2 = 1 + \frac{9}{2d}.
\end{equation} 
The above value depends on $d$, which is directly connected with the amount of GHZ entanglement. In the limit of infinite entanglement ($d\rightarrow \infty$) the value of $B_2$ goes to one, therefore the necessary condition for being a composite boson is met. In the Appendix \ref{app:comp} we also argue why in the limit of infinite entanglement $\frac{B_N}{B_{N-1}} \rightarrow 1$ should hold for all $N$.

Next, imagine that a single boson is removed from our composite particles. The state $|\mathbf{1}\rangle$ becomes $\rho_{\mathbf{1}}=\rho_2^{tri}$ and the state $|\mathbf{2}\rangle$ becomes 
\begin{eqnarray}
& &\rho_{\mathbf{2}} = \frac{1}{2\tilde{B}_2(4d)^2}\sum_{i,j =1}^{2d} a_i^{\dagger 2}a_j^{\dagger 2}|\mathbf{0}\rangle\langle\mathbf{0}| a_j^2 a_i^2 , 
\end{eqnarray}
where $\tilde{B}_2$ is a new normalization constant. This time we get
\begin{equation}
\tilde{B}_2 =  \frac{1}{2} + \frac{1}{2d},
\end{equation}
hence in the limit $d\rightarrow \infty$ the value of $\tilde{B}_2$ goes to $\frac{1}{2}$ and does not meet the necessary condition for describing a composite boson. As a result, removing a single boson from the multi-bosonic  system described by $|tri_0\rangle$ makes the whole system lose its collective bosonic property. This is yet another manifestation of this state's Borromean nature.




\section{Discussion}

In this work we have investigated a multipartite quantum walk with Grover-like interaction. Despite its simplicity, the model exhibits complex dynamics. Our main result consists in showcasing the existence of three- and four-partite Borromean states -- bound states that fall apart when particles are removed from the system. Interestingly, their formation requires genuine  multipartite entanglement of the GHZ type. 

These states can also be interpreted in terms of composite particles. From this perspective, their Borromean character manifests itself in the fact that the composite particle consisting of three and four bosons loses its bosonic character whenever one of its elementary particles is removed. Thus, our work demonstrates the crucial role of the structure of multipartite entanglement between internal degrees of freedom of particles in the creation of  Borromean composite bosons.

Moreover,  our bound states are also the eigenstates of the evolution operator resulting from replacing the identities in the definition (\ref{C}) with any qubit operators (for instance Hadamard operators commonly used in quantum walks). However, it is not obvious if these states remain Borromean, as the new evolution may allow the particles to reunite after splitting.  Nevertheless, numerical simulations suggest that this is not likely and the whole compound tends to fall apart when a single particle is removed. 

Finally, we believe that it could be worthwhile to study our model with a different choice of interaction Eq. (\ref{grov}). In particular, it remains an open problem whether there are interactions that  lead to arbitrary Brunnian states. It would also be interesting to look for relevant interactions that could be implemented experimentally.


\section{Acknowledgements}

{\it Acknowledgements.} This work is supported by the Ministry of Science and Higher Education in Poland (science funding scheme 2016-2017 project no. 0415/IP3/2016/74). We acknowledge partial support by the Foundation for Polish Science (IRAP project, ICTQT, contract no. 2018/MAB/5, co-financed by EU within Smart Growth Operational Programme). M.K. acknowledges the Grant No. 2017/27/N/ST2/01858 from the National Science Center in Poland and the support by the Foundation for Polish Science (FNP) through the START scholarship.


\appendix

\section{Discrete-time quantum walks}
\label{app:dtqw}

DTQWs describe an evolution of a quantum particle in a discrete space-time \cite{Aharonov,Meyer}. The locality condition, i.e., a single step of the evolution cannot take the particle farther than a neighbouring location, and the unitarity of the evolution imply that the particle has to be described by its position and an additional degree of freedom. This degree of freedom is known as the coin. The coin is needed because translations from two different positions, corresponding to orthogonal position states, may lead to the same location. To keep the orthogonality between the states with overlapping positions one imposes the orthogonality on the coin subspace.

Here we consider one-dimensional DTQWs for which the coin is a two-level system. The state of one particle is described by
\begin{equation}
|\psi_t\rangle = \sum_{x}\sum_c\alpha_{x,c}|x,c\rangle,
\end{equation}
where $x\in{\mathbb Z}$ and $c\in\{\leftarrow,\rightarrow\}$. A single step of the evolution is given by
\begin{equation}
|\psi_{t+1}\rangle = S(\openone\otimes C)|\psi_t\rangle.      
\end{equation}
In the above $S$ is a conditional translation operator
\begin{equation}
S|x,\rightarrow\rangle = |x+1,\rightarrow\rangle,~~S|x,\leftarrow\rangle = |x-1,\leftarrow\rangle    
\end{equation}
and $C$ is a unitary coin "toss" operator that transform the coin states. Usually this operator is chosen to be a Hadamard matrix
\begin{eqnarray}
H|\rightarrow\rangle &=& \frac{1}{\sqrt{2}}(|\rightarrow\rangle + |\leftarrow\rangle),\\
H|\leftarrow\rangle &=& \frac{1}{\sqrt{2}}(|\rightarrow\rangle - |\leftarrow\rangle),   
\end{eqnarray}
however it can be an arbitrary $2\times 2$ unitary matrix. Here we choose it to be the identity matrix, apart from the situations in which two particles meet and interact.

DTQWs spread faster than classical random walks, quadratically faster on translational invariant lattices. That is why they are used in quantum algorithms to achieve better than classical computational efficiency. In addition, they are known to be capable of simulating many physical systems. For more information we refer the reader to the following review papers \cite{Review1,Review2,Review3,Review4}. 


\section{Derivation of Borromean bound states}
\label{app:borro}

As argued in the main text, to find Borromean bound states we need to investigate the condition  
\begin{equation}\label{cn2}
\left|\langle GHZ_N|P_k\left(\prod_{j<l}G_{jl}(\varphi)\right)|GHZ_N\rangle\right| = 1,
\end{equation}
where  
\begin{equation}|GHZ_N\rangle=\beta\,|{\rightarrow},\ldots,{\rightarrow}\rangle + \gamma\,|{\leftarrow},
\ldots,{\leftarrow}\rangle,
\end{equation}
 $\prod_{j<l}G_{jl}(\varphi)$ is a sequence of generalized Grover operators for all $\frac{N(N-1)}{2}$ pairs of particles ($j,l=1,\ldots,N$) and
\begin{eqnarray}
P_k &=& e^{-i\frac{2\pi k}{d}}|{\rightarrow},\ldots,{\rightarrow}\rangle\langle {\rightarrow},\ldots,{\rightarrow}| \nonumber \\
&+& e^{i\frac{2\pi k}{d}}|{\leftarrow},\ldots,{\leftarrow}\rangle\langle {\leftarrow},\ldots,{\leftarrow} |.
\end{eqnarray}

First, we introduce a useful notation 
\begin{equation}\label{bbasis}
|1\rangle = \frac{|{\rightarrow}\rangle + |{\leftarrow}\rangle}{\sqrt{2}},~~|0\rangle = \frac{|{\rightarrow}\rangle - |{\leftarrow}\rangle}{\sqrt{2}}.
\end{equation}
Next, consider $2^N$ states which we label by binary sequences
\begin{equation}
|\pmb{\mathrm{\alpha}}\rangle=|\alpha_1,\alpha_2,\ldots,\alpha_N\rangle,
\end{equation}
where $\alpha_i = 0,1$. Every sequence corresponds to a tensor product of $N$ states, each being either $|0\rangle$ or $|1\rangle$, see (\ref{bbasis}). It is easy to verify that
\begin{equation}
G_{jk}(\varphi)|\pmb{\mathrm{\alpha}}\rangle=e^{i (\alpha_j\alpha_k) \varphi}|\pmb{\mathrm{\alpha}}\rangle.
\end{equation}
Thus
\begin{equation}
\prod_{j< k}G_{jk}(\varphi)|\pmb{\mathrm{\alpha}}\rangle=\exp{[i\varphi\sum_{j< k}\alpha_j\alpha_k ]}|\pmb{\mathrm{\alpha}}\rangle = e^{in_l\varphi}|\pmb{\mathrm{\alpha}}\rangle,
\end{equation}
where $n_l=\frac{(N-l)(N-l-1)}{2}$ and $l$ is the total number of zeros in the bit string $\pmb{\mathrm{\alpha}}$.

Then, we rewrite the $|GHZ_N\rangle$ state as 
\begin{equation}
|GHZ_N\rangle=\sum_{l=0}^{N} \eta_l |l\rangle,
\end{equation}
where $|l\rangle$ is a normalized even superposition of ${\mathcal{N}}_l = \frac{N!}{(N-l)!l!}$ states corresponding to all bit strings of length $N$ with exactly $l$ zeros and
\begin{equation}
\eta_l=\frac{\sqrt{{\mathcal{N}}_l}\left(\beta-\gamma(-1)^{l}\right)}{2^{\frac{N}{2}}}
\end{equation}
is the corresponding probability amplitude. This means that
\begin{equation}\label{GHZexp}
\prod_{j<k}G_{jk}(\varphi)|GHZ_N\rangle=\sum_{l=0}^{N} \eta_l e^{i n_l \varphi}|l\rangle.
\end{equation}
It is also useful to introduce
\begin{eqnarray}
\langle GHZ_{N}^{(k,\varphi)}|&=& \langle GHZ_N| P_k \left(\prod_{j<l}G_{jl}(\varphi)\right) \nonumber\\
&=&\sum_{l=0}^{N} \eta_{l,k}^{\ast} e^{i n_l \varphi} \langle l|,
\end{eqnarray}
where
\begin{equation}
\eta_{l,k}^{\ast}=\frac{\sqrt{{\mathcal{N}}_l}\left(e^{-i\frac{2\pi k}{d}}\beta^{\ast}-e^{i\frac{2\pi k}{d}}\gamma^{\ast}(-1)^{l}\right)}{2^{\frac{N}{2}}}.
\end{equation}

The equation (\ref{cn2}) takes form
\begin{equation}\label{cn3}
\left|  \langle GHZ_{N}^{(k,\varphi)}|GHZ_N\rangle \right|=1.
\end{equation}
It is easy to see that
\begin{eqnarray}\label{expl}
&&\left|  \langle GHZ_{N}^{(k,\varphi)}|GHZ_N\rangle \right|\leq\sum_{l=0}^{N}| \eta_{l,k}^{\ast}\eta| \nonumber\\
&&=\left||\beta|^2 e^{-i\frac{2\pi k}{d}}+|\gamma|^2 e^{i\frac{2\pi k}{d}}\right|\leq1.
\end{eqnarray}
Combining (\ref{cn3}) and (\ref{expl}) we see that $e^{i n_{l} \varphi}$ needs to be the same for all $l$ for which $ \eta_{l,k}^{\ast}\eta_l$ is nonzero and $k\in\{0,d/2\}$. These conditions are very restrictive. Notice that $ \eta_{l,k}^{\ast}\eta_l$ can be zero only if  $\beta =\pm \gamma$. The two situations correspond to superpositions of bit strings with only even or odd number of ones.  Moreover, since $n_N = n_{N-1}=0$, we must have $e^{i n_{l} \varphi}=1$ for all $l$ for which $\eta_l$ is nonzero. The above implies that (\ref{cn3}) can be satisfied only for $N=2,3,4$. More precisely, (\ref{cn3}) is satisfied if $k=0$ or $k=d/2$ and by 
\begin{equation}
|GHZ_2^-\rangle = \frac{1}{\sqrt{2}}(|\rightarrow,\rightarrow\rangle - |\leftarrow,\leftarrow\rangle)
\end{equation}
for an arbitrary $\varphi$, 
\begin{equation}
|GHZ_3^+\rangle = \frac{1}{\sqrt{2}}(|\rightarrow,\rightarrow,\rightarrow\rangle + |\leftarrow,\leftarrow,\leftarrow\rangle)
\end{equation}
for $\varphi = \frac{2\pi}{3}$, 
\begin{equation}
|GHZ_4^-\rangle = \frac{1}{\sqrt{2}}(|\rightarrow,\rightarrow,\rightarrow,\rightarrow\rangle - |\leftarrow,\leftarrow,\leftarrow,\leftarrow\rangle)
\end{equation}
for $\varphi = \frac{2\pi}{3}$.


\section{Evolution of two particles}
\label{app:two}

Here we will discuss the dynamics of two particles under the projected evolution operator $V_2 = \Pi_2 U_2$.
Let us start by making the following ansatz
\begin{eqnarray}
|k\rangle_2 &=& \frac{1}{\sqrt{d}}\left(\sum_{x=0}^{d-1}e^{i\frac{2\pi}{d}kx}|x,x\rangle\right) \nonumber \\ &\otimes& (a_k |\rightarrow,\rightarrow\rangle + b_k|\leftarrow,\leftarrow\rangle),
\end{eqnarray}
where $k=0,1,\ldots,d-1$. It is easy to see that due to the translational symmetry of the position state the eigenvalue problem $V_2|k\rangle_2 = \lambda_k |k\rangle_2$ reduces to the following $2\times 2$ problem
\begin{equation}\label{ev1}
\begin{pmatrix} he^{-i\frac{2\pi}{d}k} & ge^{-i\frac{2\pi}{d}k} \\ ge^{i\frac{2\pi}{d}k} & he^{i\frac{2\pi}{d}k} \end{pmatrix} \begin{pmatrix} a_k \\ b_k \end{pmatrix} = \lambda_k \begin{pmatrix} a_k \\ b_k \end{pmatrix},
\end{equation} 
where $g = \frac{1}{8}(-3+i\sqrt{3})$ and $h = 1+g$. The corresponding eigenvalues are
\begin{equation}\label{ev2}
\lambda_{k,\pm} = h\cos\left(\frac{2\pi}{d}k\right) \pm \sqrt{g^2-h^2\sin^2\left(\frac{2\pi}{d}k\right)}.
\end{equation}
There are only two eigenvalues with modulus one, namely $\lambda_{0,-}=1$ and $\lambda_{d/2,+}=-1$. They correspond exactly to
\begin{equation}
|dim_r\rangle = \frac{1}{\sqrt{d}} \left( \sum_{x=0}^{d-1} (-1)^{rx}|x,x\rangle \right) \otimes |GHZ_2^-\rangle
\end{equation}
with $r=0,1$. 

\begin{figure}[t]
\centering
\includegraphics[scale=1]{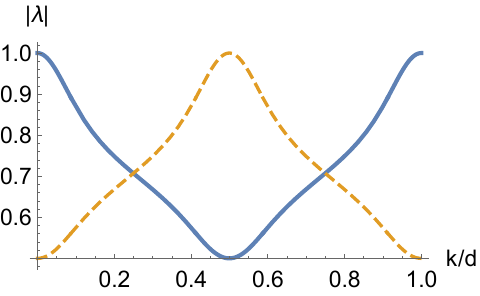}
\caption{Plot of the norm of $\lambda_{k,+}$ (solid) and $\lambda_{k,-}$ (dashed).}
\label{fig3}
\vspace{-1mm}
\end{figure}

Although $|dim_0\rangle$ and $|dim_1\rangle$ are the only cases that will not split at all, one can also ask how long the particles would remain together in other states. This depends on the norms of all the eigenvalues (\ref{ev2}) presented in Fig. (\ref{fig3}) in the limit  $d\rightarrow\infty$. The shape of these distributions implies that for a large number of steps $t$ there will be only a finite number of eigenvectors whose contributions cannot be neglected. In particular, this means that the state
\begin{equation}
\rho^{tri}_2 = \frac{1}{2d}\sum_{x=0}^{d-1}\sum_{c={\leftarrow},{\rightarrow}} |x,x,c,c\rangle\langle x,x,c,c|
\end{equation}
vanishes as $t$ and $d$ go to infinity. In Fig. (2) we present numerically obtained probabilities showcasing the disappearance of the bound state $\rho^{tri}_2$ during the quantum walk on a finite lattice.

\begin{figure}[h]
\centering
\includegraphics[scale=0.7]{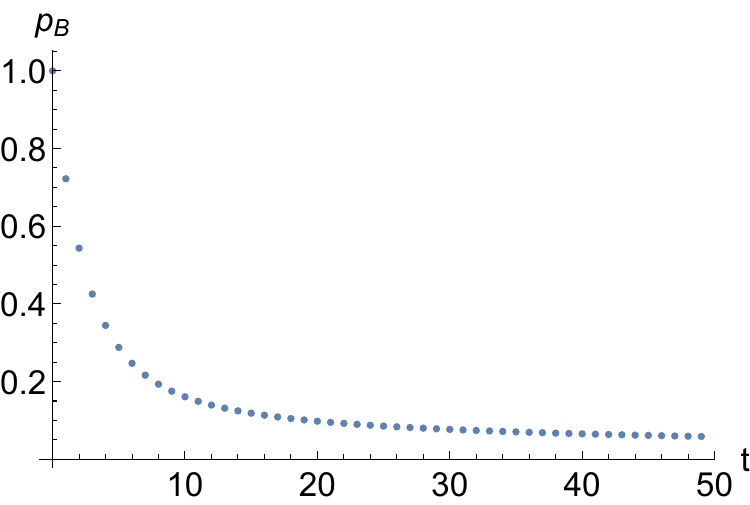}
\caption{Probability $p_B$ that the state $\rho^{tri}_2$ does not fall apart after $t$ steps of the walk on $d=100$ vertices.}
\label{figrepl}
\vspace{-1mm}
\end{figure}


\section{Composite bosons and the normalization factor}
\label{app:comp}

The idea that entanglement is responsible for composite bosonic behaviour was introduced by Law in \cite{Law}, where a bipartite case was discussed. In particular, Law showed that composite bosonic particles, each made of two bosons (or two fermions), reproduce the bosonic ladder rules, i.e.,  $a^{\dagger}|N\rangle = \sqrt{N+1}|N+1\rangle$ and $a|N\rangle = \sqrt{N}|N-1\rangle$, if the constituent elementary particles are strongly entangled. The stronger the entanglement, the better the bosonic behaviour of the composites.  

In our case we consider a composite boson made of three elementary bosons in a Borromean state. The composite particle is described by a creation operator
\begin{equation}
b^{\dagger} = \frac{1}{\sqrt{3!}}\frac{1}{\sqrt{2d}}\sum_{i=1}^{2d} a^{\dagger 3}_i,
\end{equation}
where for simplicity we introduced a convention $a^{\dagger}_{x,c}=a^{\dagger}_i$ and $i$ is an index that runs through all $\{x,c\}$ pairs. We notice that
\begin{equation}
|N\rangle = \frac{b^{\dagger N}}{\sqrt{B_N N!}}|0\rangle,    
\end{equation}
where the value $B_N$ describes the deviation from perfect bosonic behaviour.

The first goal is to show that in the limit of strong entanglement the value
\begin{equation}
B_2 = \frac{1}{2}\langle \mathbf{0} |b^2 b^{\dagger 2}|\mathbf{0}\rangle, 
\end{equation}
approaches one. Hence, we need to evaluate the norm of
\begin{eqnarray}
b^{\dagger 2}|\mathbf{0}\rangle  =\frac{1}{(3!)2d} \sum_{i,j=1}^{2d}a^{\dagger 3}_{i} a^{\dagger 3}_{j}|\mathbf{0}\rangle.
\end{eqnarray}
Since $a_i^{\dagger}$ and $a_j^{\dagger}$ commute, the above can be rewritten as
\begin{eqnarray}
b^{\dagger 2}|\mathbf{0}\rangle  &=& \frac{1}{(3!)2d} \left(2\sum_{i<j}a^{\dagger 3}_{i} a^{\dagger 3}_{j} + \sum_{i=1}^{2d}a^{\dagger 6}_{i} \right)|\mathbf{0}\rangle
\nonumber\\
&=& \frac{1}{d} \sum_{i<j}|3_i,3_j\rangle +  \frac{\sqrt{6!}}{(3!)2d}\sum_{i=1}^{2d}|6_i\rangle,
\end{eqnarray}
where $|3_i,3_j\rangle$ is the Fock state representing three particles in mode $i$ and three particles in mode $j$, whereas $|6_i\rangle$ represents six particles in mode $i$. The corresponding norm is
\begin{equation}
2B_2 = \frac{2d-1}{d} + \frac{10}{d} = 2 + \frac{9}{d}.
\end{equation} 

Let us also provide an argument why $\frac{B_N}{B_{N-1}} \rightarrow 1$ should hold for all $N$. Notice that for $d\gg N$ the vector $b^{\dagger N}|\mathbf{0}\rangle$ is dominated by the terms in which the triples of elementary bosons occupy different modes. In this case
\begin{equation}
b^{\dagger N}|\mathbf{0}\rangle \approx \frac{N!}{(2d)^{N/2}} \sum_{i_1<\ldots <i_N} |3_{i1},\ldots,3_{i_N}\rangle.
\end{equation}
The corresponding norm is
\begin{equation}
N! B_N \approx \frac{(N!)^2}{(2d)^{N}} \frac{(2d)!}{N!(2d-N)!}.
\end{equation}
Therefore,
\begin{equation}
\frac{B_N}{B_{N-1}} \approx  \frac{2d-N+1}{2d} \approx 1. 
\end{equation}

Finally, let us find the normalization constant $\tilde{B}_2$ of the state
\begin{eqnarray}
& &\rho_{\mathbf{2}} = \frac{1}{2\tilde{B}_2(4d)^2}\sum_{i,j =1}^{2d} a_i^{\dagger 2}a_j^{\dagger 2}|\mathbf{0}\rangle\langle\mathbf{0}| a_j^2 a_i^2.
\end{eqnarray}
 Notice that
\begin{eqnarray}
& &2\tilde{B}_2\rho_{\mathbf{2}} = \frac{1}{(4d)^2}\sum_{i,j =1}^{2d} a_i^{\dagger 2}a_j^{\dagger 2}|\mathbf{0}\rangle\langle\mathbf{0}| a_j^2 a_i^2  \\
& &= \frac{2}{(4d)^2}\sum_{i<j}^{2d} a_i^{\dagger 2}a_j^{\dagger 2}|\mathbf{0}\rangle\langle\mathbf{0}| a_j^2 a_i^2 + \frac{1}{(4d)^2}\sum_{i =1}^{2d} a_i^{\dagger 4}|\mathbf{0}\rangle\langle\mathbf{0}| a_i^4, \nonumber 
\end{eqnarray}
Then, it is straightforward to evaluate
\begin{equation}
2\tilde{B}_2 = \frac{1}{(4d)^2} \left(8d(2d-1) + (4!)d \right) = 1 + \frac{1}{d}.
\end{equation}

\end{document}